\documentclass[11pt]{article}
\usepackage{amssymb}
\usepackage{color,graphicx}
\usepackage[english]{babel}
\usepackage{epsf}
\title{\bf On the Casimir energy of the electromagnetic  field in the dispersive
and absorptive medium}
\author{M.A.Braun,\\
 University of S.Petersburg, 198504 S.Petersburg, Russia}
\def\beq{\begin{equation}}
\def\eeq{\end{equation}}
\def\bae{{\bf A}}
\def\bee{{\bf E}}
\def\bbe{{\bf B}}
\def\bx{{\bf X}}
\def\bk{{\bf k}}
\def\bxp{{\bf X}_\perp}
\def\by{{\bf Y}_\omega}
\def\byp{{\bf Y}_{\omega\perp}}

\def\cl{{\cal L}}
\def\an{a_n}
\def\and{a^\dagger_n}
\def\bo{b_{n,\omega}}

\def\bop{b_{n,\omega'}}
\def\bopd{b_{n,\omega'}^\dagger}
\def\io{\int_0^\infty d\omega}
\def\bb{B_{n,\omega}}
\def\bbd{B_{n,\omega}^\dagger}
\def\bbp{B_{n,\omega'}}
\def\bbpd{B_{n,\omega'}^\dagger}
\def\o{\omega}
\def\ep{\epsilon}
\def\im{{\rm Im}}

\begin{document}
\maketitle
\medskip
{\bf Abstract.}

The microscopic theory of the Casimir effect in the dielectric
is studied in the framework when  absorption is realized
via a reservoir modeled by a set of oscillators with continuously
distributed frquencies with the aim to see if the effects depend on
the form of interaction with the reservoir. A simple case of the
one-dimensional dielectric between two metallic plates is considered.
Two possible models are investigated, the direct interaction of the
electromagnetic field with the reservoir and indirect interaction via
an intermediate oscillator imitating the atom. It is found that with
the same dielectric constant the Casimir effect is different in these
two cases, which implies that in the second model it cannot be entirely
expressed via the dielectric constant as in the well-known Lifshitz
formula.

\section{Introduction}
The consistent quantum-mechanical treatment of the Casimir forces  in the
dispersive and absorbing dielectric requires inclusion of the absorbing
medium as an independent dynamical system.
This problem has attracted attention since long ago.
A formalism allowing to consider absorption of the electromagnetic
field in the medium in the microscopic approach
was developed by B.Huttner and S.M.Barnett ~\cite{hutbar}.
It consists in modeling the medium as a set of oscillators with
continuously distributed frequencies.
Interaction with the medium
leads to absorption of the electromagnetic field and its exclusion as
an independent dynamical
variable. The resulting system is completely described by a set of
effective oscillators, which also have  continuously distributed
frequencies. Quantum field excitations are expressed via the ones  of
these effective oscillators ("polaritons").

Within this or similar picture introduced explicitly or assumed implicitly
derivation of the Casimir energy can be done using the macroscopic
expression for it and interpreting the electromagnetic field as a quantum
operator satisfying the Heisenberg equations in which the influence of the
medium appears as a "quantum noise" leading to absorption. Taking the average
in the ground state one obtains the Casimir energy of the field in the
presence of the medium. This or similar approach was presented in various
publications ~\cite{mochan,lombardo, ram1, interavia}. The problem of such
a treatment is taking fully into account  the interaction energy between the
field and medium and  determination of the ground state, which may change
with this interaction.

The consistent treatment requires to determine the ground state of the
total Hamiltonian and take the average in this state. Such a program
was accomplished in ~\cite{ram2, phil}, where however a simplified version
of the original Huttner-Barnett (HB) model was studied. In the
HB model the electromagnetic  field is assumed to
interact directly only
with atoms in the dielectric and the latter to dissipate afterwards their
energy by their interaction with the medium. In the simplified version
the field directly interacts with the medium without the intermediary atom
(the direct (D) model).
Additionally, to avoid the change of the ground state with interaction,
the authors of ~\cite{ram2}
chose the interaction in an unnatural way to depend
separately on creation and annihilation operators  for the field and
medium variables. The final formulas of both calculations are
somewhat different. They both have the form of the classical expression
for the energy in which the (real)
refractive index $n=\sqrt{\epsilon}$, where $\epsilon$ is the
dielectric constant, is to be substituted by the full complex $n$
in ~\cite{phil} and by its real part in ~\cite{ram2}.

With all this the Casimir energy in the initial HB model
has never been calculated consistently, that is
as the ground state energy of the total Hamiltonian taking into account that
the ground state itself changes with interaction. This calculation
occupies the main part of this paper.

Note that this problem is of a wider scope. Modeling the absorbing medium
and its interaction with the field it is important to know if the final
results for, say, the Casimir energy depend on different properties of a
particular model or this dependence is wholly concentrated in the way
the field propagates in the medium, that is in the complex and
frequency-dependent
electromagnetic constants $\epsilon$ and $\mu$ of the medium.
A remarkable result found  in papers ~\cite{ram2} and \cite{phil}
(although different)  is that the
Casimir energy can be expressed entirely in terms of the frequency
dependent complex dielectric and magnetoelectric constants thus
linking this approach with the standard macroscopic Lifshitz formula
~\cite{lifshitz}. The influence of the medium was found to be
implicitly included into the properties of the two constants.
The question is whether this result is general or restricted to
specific forms of the model.

In this paper we show that for models in which the field interacts
directly with the medium, which are generalizations of the simple
models of  ~\cite{ram2} and \cite{phil}, this result remains valid.
However for the more complcated original HB model it does not.
Due to complexity of the
derivation in the HB model we obtained this result only for a
particulary simple case of one dimensional electromagnetic field between
two metallic plates.
Moreover even with this simplification we could
find it only numerically and with a specific form of the atom-medium
interaction. With this interaction in the HB model we  found  both the dielectric
constant $\epsilon(\o)$ and Casimir energy.
Then we compared this energy with the one calculated in the simple
model of T.Philbin ~\cite{phil} using the same $\epsilon(\o)$.
If the energy is wholly determined by the dielectric constant the result
should also be the same. However in fact
the resulting Casimir energy proved to be very different indicating that
the Casimir energy depends not only on $\epsilon(\o)$ but on the details
of the model.
So the answer to the question
whether the Casimir force is uniquely determined by the dielecric (and
in all probability magnetoelectric) constant seems to be  negative.

This also means that at least in this framework the experimental study
of the Casimir force may give some information about the dynamical
mechanism behind dispersion and absorption in the medium.

\section{Simple model (D model)}
We start with a simple model introduced by T.Philbin for the quantization
of the electromagnetic field in the dispersive and absorptive medium.
To further simplify we restrict ourselves with a homogenious dielectric as
the medium, represented microscopically by field $\by$ with continuously
distributed frequencies. The Lagrangian density splits into three parts
$
\cl= \cl_e+ \cl_r+\cl_i.
$
Here $\cl_e$ is the Lagrangian density of the electromagnetic field
\beq
\cl_e=\frac{1}{2}
\Big({\dot \bee}^2-\bbe^2\Big),
\eeq
with
\beq
\bee=-\dot{\bae}-\nabla\phi,\ \ \bbe=\nabla\times\bae.
\eeq
$\cl_r$ is the Lagrangian density of the "reservoir",
\beq
\cl_r=\frac{1}{2}\int_0^\infty d\o
\Big({\dot \by}^2-\o^2\bx^2\Big)
\eeq
and finally $\cl_i$ is the Lagrangian density for the interaction,
which we take following ~\cite{phil}
\beq
\cl_i=
-\int_o^\infty d\o v(\o)\bae\dot{\by},
\eeq
where $v^2(\omega)$ is a  square integrable function which can be
analitically continued to negative $\omega$ as an even function.
In contrast with T.Philbin we do not introduce interaction with the
scalar potential $\phi$ nor with the magnetic field assuming that
the medium is magnetically neutral.
In the Coulomb gauge the dynamical part of the Lagrangian density becomes
expressed effectively via transverse fields
\beq
\cl_\perp=
\frac{1}{2}
\Big(\dot {\bae}^2-\bbe^2\Big)+
+
\frac{1}{2}\int_0^\infty d\o
\Big(\dot {\by}_\perp^2-\o^2\byp^2\Big)
-\int_o^\infty d\o v(\o)\bae\cdot\dot{\byp}.
\label{lag1}
\eeq

Passing to the momentum space the transverse Lagrangian becomes a sum over
two polarizations $\lambda=1,2$ with
\beq
L_\lambda=\int d^3k
\Big\{\frac{1}{2}\Big(|\dot{A}_\lambda|^2-k^2|A_\lambda|^2+
+
\int_0^\infty d\o |\dot{Y}_{\o,\lambda}|^2-\o^2|Y_{\o,\lambda}|^2\Big)
-
\int_0^\infty d\o v(\o)X_\lambda^*\dot{Y_{\o,\lambda}}\Big\}.
\eeq
Each polarization is treated similarly and in following subindex $\lambda$
will be suppressed.

At this point we introduce our final simplification passing to the
one-dimensional space $0<x<a$ and imposing the boundary conditions for
two metallic plates
\[ A(x=0)=A(x=a)=Y_\o(x=0)=Y_\o(x=a)=0.\]
For the electromagnetic field this means that the plates are ideal
reflectors, all dissipation coming only from the dielectric between
the plates. As to the reservoir field $Y$, the boundary
conditions may of course be taken in different ways
but having in mind our restricted aim presented in the Introduction we
choose the simplest and most convenient form.

Then the lagrangian becomes a sum over discrete values of momentum
\[ k_n=\frac{\pi n}{a},\ \ n=1,2,...\]
and fields can be expanded as
\[A(x)=\sqrt{\frac{2}{a}}A_n\sin(k_nx),\ \
Y_\o(x)=\sqrt{\frac{2}{a}}Y_{n\o}\sin(k_nx).
\]

Quantization then follows in the standard manner, introducing the conjugate
fields, $\pi(x)$ and $\Pi_\o(x)$ for $A(x)$ and $Y_\o(x)$ respectively,
and imposing the standard commutation relations.
In terms of creation and annihilation operators $a_n,a_n^\dagger$ for the
electromagnetic field and $b_{n,\o},b^{\dagger}_{n,\o}$ for the medium
\beq
A_n=\frac{1}{\sqrt{2\omega_n}}(a_n+a^\dagger_n),\ \
\pi_n=-i\sqrt{\frac{\omega_n}{2}}(a_n-a^\dagger_n),
\label{phin}
\eeq
and
\beq
Y_{n,\omega}=\frac{i}{\sqrt{2\omega}}(b_{n,\omega}-b_{n,\omega}^\dagger),
\ \
\Pi_{n,\omega}=\sqrt{\frac{\omega}{2}}(b_{n,\omega}+b_{n,\omega}^\dagger)
\label{yin}.
\eeq

One finds two equivalent expressions for the total Hamiltonian.
In terms of fields and their time derivatives
\beq
H=\frac{1}{2}\sum_n\Big\{{\dot{A}}^2_n+k_n^2A_n^2
+\int_0^\infty
d\omega({\dot{Y}}^2_{n,\omega}+\omega^2 Y^2_{n,\omega}\Big)\Big\}
\label{h1}
\eeq
or in terms of fields and their conjugates
\beq
H=\frac{1}{2}\sum_n\Big(\pi^2_n+k_{1n}^2A_n^2
+\int_0^\infty
d\omega(\Pi^2_{n,\omega}+\omega^2 Y^2_{n,\omega}\Big)+
 \sum_n\int_0^\infty d\omega v(\omega)A_n \Pi_{n,\omega}\Big)
 \label{h2},
\eeq
where
\beq
k_{1n}^2\equiv k_n^2+\mu^2=k_n^2+\int_0^\infty d\omega v^2(\omega).
\label{omn}
\eeq

\section{Fano diagonalization, ground state energy and Casimir energy}

In terms of creation and annihilation operators the Hamiltonian has the form
\[
H=\frac{1}{2}\sum_n\Big[k_{1n}\{a^\dagger_n,a_n\}
+\int_0^\infty d\o \o\{b_{n\o}^\dagger,b_{n\o}\}\]\beq+\frac{1}{2}
\int_0^\infty d\o (\o)[a^\dagger_n+a_n][V_n(\o)b_{n\o}^\dagger
+V^*(\o)b_{n\o}]\Big].
\label{hgen}
\eeq
Here
\[V_n(\o)=\sqrt{\frac{\o}{k_{1n}}}v(\o).\]
It can be demonstrated that the  Hamiltonian $H$ can be diagonalized
introducing new field variables ~\cite{hutbar}
\beq
B_{n,\omega}=\alpha_{0n}(\o)\an+\beta_{0n}(\o)\and
+\int_0^\infty d\omega'\Big[\alpha_{1n}(\o,\o')\bop+\beta_{1n}
(\o,\o')\bopd\Big],
\label{newold}
\eeq
which satisfy the commutation relations
\beq
[B_{n,\o},B^\dagger_{n',\o'}]=\delta_{nn'}\delta(\omega-\omega'),\ \
[B_{n,\o},B_{n',\o}]=0.
\label{comrel}
\eeq
These commutation relations together with the requirement that
\beq
[B_{n,\o},H]=\omega B_{n,\o}
\label{hrel}
\eeq
uniquely define the coefficients
$\alpha_{0n},\ \beta_{0n},\ \alpha_{1n},\ \beta_{1n}$
(see \cite{hutbar}):
\beq
\alpha_0(\o)=-\Big(\frac{\o+\o_1}{2}\Big)V_n(\o)P_n^*(\o),\ \
\beta_0(\o)=-\Big(\frac{\o-\o_1}{2}\Big)V_n(\o)P_n^*(\o),
\label{albe}
\eeq
\beq
\alpha_1(\o,\o')=\delta(\o-\o')-\Big(\frac{\o_1}{2}\Big)
\Big(\frac{V^*_n(\o') V_n(\o)}{\o-\o'-i0}\Big)P_n^*(\o),
\label{al1}
\eeq
\beq
\beta_1(\o,\o')=-\Big(\frac{\o_1}{2}\Big)
\Big(\frac{V_n(\o') V_n(\o)}{\o+\o'}\Big)P^*_n(\o),
\label{be1}
\eeq
where
\beq
P_n(\o)=\frac{1}{k^2-\ep(\o)\o^2}
\label{prop1}
\eeq
is the propagator of the electromagnetic field in the medium
and
\beq
\ep(\o)=1+\frac{k_{1n}}{\o}\int_{-\infty}^{\infty}
\frac{d\o'}{\o'}\frac{|V_n(\o')|^2}{\o'-o-i0}
\label{ep}
\eeq
is the dielectric constant.

In terms of the new variables Hamiltonian $H$ has
the form
\beq
H=\sum_n\int_0^\infty d\omega \omega B^\dagger_{n,\omega}B_{n,\omega}+E(a).
\label{hviabb}
\eeq
Here $E(a)$ is a constant which obviously has the meaning of the ground
state energy of the system.

Assuming that the new operators $\bb$ and $\bbd$
form a complete set one can invert relations (\ref{newold}) and its
conjugate and express the initial operators as  linear superpositions
of the new ones. Comparing the commutation relations between the old and
new operators written in terms of the old and new operators one obtains
\[
\an=\int_0^\infty d\omega\Big[\alpha_{0n}^*(\omega)\bb-\beta_{0n}(\omega)
\bbd\Big],
\]\beq
\bo=\int_0^\infty d\omega'\Big[\alpha_{1n}^*(\omega',\omega)\bbp-
\beta_{1n}(\omega',\omega)\bbpd\Big].
\label{oldnew}
\eeq

This procedure is consistent provided a certain consistency relation
is satisfied ~\cite{hutbar}:
\beq
\io v|(\o)|^2<k_{1n},
\label{consist}
\eeq
which in our case is true due to (\ref{omn}).

Expressing in the expression for the Hamiltonian the old operators
$A_n$, $Y_{n,\o}$ their conjugates and time derivatives via the new
ones and taking the average in the ground state determined by the condition
$\bb|0>=0$ one can find the ground state energy $E_0$.
One can use both forms (\ref{h1}) and (\ref{h2}) for this purpose.
With (\ref{h1}) used in ~\cite{phil} the resulting formulas are
simpler and directly expressed via the dielectric constant $\ep(\o)$.
However, as a price, in the course of the derivation one has to disentangle
finite contributions from  initially singular expressions. Adjusting the
results of ~\cite{phil} to our one-dimensional case
one finds the ground state energy
\beq
E_0=\frac{1}{2\pi}
\sum_n\io\im\,\Big(\o^2\frac{d}{d\o}[\o\ep(\o)]+k_n^2\Big)P_n(\o).
\label{e01}
\eeq
For calculations one standardly rotates the contour to pass along the
positive imaginary axis to find
\beq
E_0=\frac{1}{2\pi}\sum_n\int_0^\infty d\xi
\Big\{k_n^2-\xi^2\Big(\frac{d}{d\o}[\o\ep(\o)]\Big)_{\o=i\xi}\Big\}P_n(i\xi)
\label{ce01}
\eeq
where the integrand is real.

To find the Casimir energy one has to subtract from $E_0$ its value for
the case when there are no plates, that is for $k$ continuosly distributed
in the interval $[0,\infty)$
\[E_{cas}=E_0(a)-\tilde{E}_0(a).\]
In $\tilde{E}_0$ the summation over $k_n$ is changed to integration
over $k$ with weight $a/\pi$, which implies in our formulas
\[\sum_n\to \int_0^\infty dn,\ \ {\rm with},\ \ k_n=\pi n/a.\]

\section{Simple generalizations.}
In this section to study the dependence of the Casimir energy and force
on the assumed model for the dispersive and absorbing medium we study
simple generalizations of the model presented before.
From the start to simplify we restrict ourselves to the same picture
of one dimensional fields between two metallic plates.

\subsection{Interaction with several $\dot{Y}$}

The absorbing medium is now modeled by a set of different
oscillators $Y_{j,\omega}(x)$ in the same
interval of coordinates $[0,a]$ with
continuously distributed frequencies:
\beq
 L_1=\frac{1}{2}\int_{0}^{a}dx
 \int_0^{\infty} d\omega\sum_{j=1}^N
 \Big(\dot {Y}_j^2-\omega^2 Y_j^2\Big).
\eeq
The interaction betwen the quantum field
and the medium can be generalized as
\beq
L_I=-\int_0^a dx\int_0^\infty d\omega A\sum_{j=1}^Nv_j \dot {Y}_j,
\label{li}
\eeq
where $v^2(j,\omega)$ are square integrable functions which can be
analitically continued to negative $\omega$ as an even function.

We make a unitary transformation to new variables $Y'_{j,\omega}(x)$
\beq
Y'_{j,\o}(x)=\sum_{l=1}^Nu_{jl}(\o)Y_{l,\o}(x)
\eeq
and take
\beq
\sum_{l=1}^Nv_l(\o)Y_{l,\o}(x)=v(\o)Y'_{1,\o}(x),\ \
v_l(\o)=v(\o)u_{1l}(\o).
\eeq
From the unitarity of the transformation we have
\beq
v(\o)=\sqrt{\sum_{l=1}^Nv_l^2(\o)}.
\eeq
In terms of new variables $L_1$ does not change but
$L_I$ becomes dependent only on $Y'_1$
\beq
L_I=-\int_0^a dx\int d\omega  v(\o)A\dot {Y}'_{1,\o},
\label{li2}
\eeq

As a result the model is completely equivalent to the one
with a single $Y$, all the additional variables of the medium
not interacting with the electromagnetic field.

\subsection{Interaction with both $\dot{Y}$ and $Y$}
Next we study a generalization to interaction with both
$\dot{Y}$ and $Y$ with the interaction Lagrangian
\beq
L_I=-\int_0^a dx\int d\omega A \Big(v_1\dot {Y}-v_2Y\Big),
\label{li1}
\eeq
To quantize we determine the conjugated variables as usual
and find the Hamiltonian in the form
$
H_e+H_Y+H_I
$
where $H_e$ and $H_Y$ for the free electromagnetic field and medium
are  the same as before but with the mass shift
\beq
\mu^2=\io v_1^2(\o).
\label{mu1}
\eeq
Remarkably it depends only on $v_1$.
The interaction is now
\beq
H_I=\int_0^adx\int_0^\infty d\o\ A\Big(v_1\Pi_\o+v_2Y_\o\Big).
\eeq

Our strategy is the same: we try to reduce this to the old model.
First we rescale variables $Y$ and $\Pi$
\beq
Y_\o=\frac{1}{\sqrt{\o}}Q_\o,\ \ \Pi_\o=\sqrt{\o}P_\o,
\eeq
so that
\beq
H_1=\int_0^a dx\int_0^\infty \o(Q_\o^2+P_\o^2).
\eeq
Next we do a canonical transformation
\beq
Q'_\o=Q_\o\cos\theta+P_\o\sin\theta,\ \
P'_\o=-Q_\o\sin\theta+P_\o\cos\theta.
\eeq
It preserves the form of $H_1$ and commutation relations between
$Q$ and $P$ which are standard.
Now we identify
\beq
v_1\Pi_\o+v_2Y_\o=\frac{v_2}{\sqrt{\o}}Q_\o+v_1\sqrt{\o}P_\o=
\tilde{v}(\o)P'_\o=\tilde{v}(\o)\Big(-Q_\o\sin\theta+P_\o\cos\theta\Big).
\eeq
Comparison gives
\beq
\tilde{v}(\o)\sin\theta=-\frac{v_2}{\sqrt{\o}},\ \
\tilde{v}(\o)\cos\theta=v_1\sqrt{\o},
\eeq
so that
\beq
\tilde{v}^2=\o v_1^2+\frac{1}{\o}v_2^2.
\eeq
Returning to the natural momenta
\beq
P'_\o=\frac{1}{\sqrt{\o}}\Pi'_\o,\ \ Q'_\o=\sqrt{\o}Y_\o
\eeq
we find that $H_1$ is the same but the interaction is now
\beq
H_I=\int_0^adx\int_0^\infty d\o  v(\o)A\Pi'_\o,
\eeq
where
\beq
v^2=\frac{1}{\o}\tilde{v}^2=v_1^2+\frac{v_2^2}{\o^2}.
\eeq

We see a problem: the mass shift depends on only $v_1$. As a result
the necesary conditions  for the consistency of the quantization
~\cite{hutbar} become violated and
the propagator develops an extra
pole on the imginary axis, which violates validity of the commutation
relations

To remedy this defect one has to include an extra
term  into the interaction to make the mass shift consistent with the
interaction $H_I$:
\beq
\Delta H_I=-\Delta L_I=\frac{1}{2}\int_0^a dx A^2\int_0^\infty
\frac{d\omega}{\omega^2}v_2^2(\o).
\eeq
The new mass-shift added to  the right-hand side of Eq. (\ref{omn})
converts the final mass into
\beq
\mu^2=\int_0^\infty d\o v^2(\o),
\eeq
which guarantees absence of extra poles of the propagator on the
complex plane.

\subsection{General case}
Now we are in the position to consider the case of $N$
oscillators in the medium with a general interaction with the field
\beq
H_I=\int_0^adx\int_0^\infty d\o\ A\sum_{j=1}^N\Big(v_j^{(1)}\Pi_j
+v_j^{(2)}Y_j\Big)
\eeq
with the mass shift
\beq
\mu^2=\sum_{j=1}^N{v_j^{(1)}}^2.
\eeq
We first canonically transform each pair $Y_j,\ \Pi_j$ to
$Y'_j\ \Pi'_j$ as before
to reduce the interaction to
\beq
H_I=\int_0^adx\int_0^\infty d\o A \sum_{j=1}^Nv^{(0)}_j\Pi'_j,
\eeq
where
\beq
{v_j^{(0)}}^2={v_j^{(1)}}^2+\frac{{v_j^{(2)}}^2}{\o^2}.
\eeq

Then we act as in the first subsection and unitary transform $Y'_j$
and $\Pi'_j$ between themselves to reduce the interaction to only with the
$\Pi''_1$
\beq
H_I=\int_0^adx\int_0^\infty d\o\ A \Pi''_1
\eeq
where now
\beq
v^2=\sum_{j=1}^N{v_j^{(0)}}^2.
\eeq

To secure absence of the poles of the propagator on the complex plane we
additionally introduce extra interaction in the form
\beq
\Delta H_I=-\Delta L_I=\frac{1}{2}\int_0^a dxA^2\int_0^\infty
\frac{d\omega}{\omega^2}\sum_j {v_j^{(2)}}^2(\o),
\eeq
so that the final mass turns out into
\beq
\mu^2=\int_0^\infty d\o v^2(\o)
\eeq
in accordance with the necessary conditions for the absence
of extra poles.
As a result, effectively this generalized model is equivalent to the old one.

\section{Further generalization: two stage scenario (the HB model)}
\subsection{The model}
In this section we introduce the picture of the medium originally
proposed by T.Huttner and S.M.Barnet in ~\cite{hutbar} in which the
electromagnetic field interacts with the absorbing medium not directly
but via an oscillator imitating the atom immersed in the medium.
The transverse Lagrangian desnsity instead of Eq. (\ref{lag1})
is taken to be
\[
\cl_\perp=
\frac{1}{2}
\Big(\dot {\bae}^2-\bbe^2\Big)+
\frac{1}{2}
\Big(\dot {\bx}^2_\perp-\o_0^2\bx^2_\perp\Big)+
\frac{1}{2}\int_0^\infty d\o
\Big(\dot {\by}_\perp^2-\o^2\byp^2\Big)\]\beq
-\alpha\bae\cdot\dot{\bxp}
-\int_o^\infty d\o v(\o)\bxp\cdot\dot{\byp}.
\eeq
with a new field $\bx$ representing the atom.
Correspondingly to our expressions for the energy new terms are
to be added corresponding to the free field $\bx$ and its interaction
with the electromagnetic field and medium.
Thus instead of Eqs. (\ref{h1}) and (\ref{h2}) in the one-dimensional case
we find two equivalent
expressions for the new Hamiltonian
\beq
H=\frac{1}{2}\sum_n\Big({\dot{A}_n}^2+k_n^2A_n^2+
\dot{X}_n^2+\o_0^2X_n^2\Big)
+\int_0^\infty
d\omega({\dot{Y}}^2_{n,\omega}+\omega^2 Y^2_{n,\omega}\Big)
\label{h21}
\eeq
or in terms of fields and their conjugates $\pi$, $q$ and $\Pi_\o$ for
$A$, $X$ and $Y_\o$ respectively
\[
H=\frac{1}{2}\sum_n\Big(\pi^2_n+k_{1n}^2A_n^2+q_n^2+\o_1X_n^2\Big)
+\int_0^\infty
d\omega\Big(\Pi^2_{n,\omega}+\omega^2 Y^2_{n,\omega}\Big)\]\beq+
 \sum_n(-\alpha A_n\dot{X}_n-
 \int_0^\infty d\omega v(\omega)X_n \Pi_{n,\omega}\Big)
 \label{h22},
\eeq
where now
\beq
k_{1n}^2=k_n^2+\alpha^2,\ \
\o_1^2=\o_0^2+\int_0^\infty d\omega v^2(\omega).
\label{omn2}
\eeq

For the following we shall need the expression for the Hamiltonial in terms
of annihilation and creation operators $a_n,a_n^\dagger$,
$b_n,b_n^\dagger$ and $b_{n\o},b_{n\o}^\dagger$ for the fields
$A$,$X$ and $Y_\o$ respectively. We have
\beq
H=\tilde{H}+E_{e0}+E_{X0}+E_{Y0}
\eeq
where
\beq
\tilde{H}=H_e+H_X+H_Y+H_{XY}+H_{eX}
\eeq
with  the free parts
\beq
H_e=\sum_nk_{1n}a^\dagger_na_n,\ \
H_x=\sum_n\o_1b_n^\dagger b_n,\ \
H_Y=\sum_n\io \o b_{n,\o}^\dagger b_{n,\o}
\label{hexy0}
\eeq
and the interaction parts
\beq
H_{XY}=\frac{1}{2}\int_0^\infty d\o V(\o)
[b^\dagger_n+b_n][b_{n,\o}^\dagger+b_{n,\o}],\ \
H_{eX}=\frac{1}{2}i\sum_n\Lambda_n[a^\dagger_n+a_n][b^\dagger_n-b_n].
\label{hexy1}
\eeq
Here
\beq
V^2(\o)=v^2(\o)\frac{\o}{\o_1},\ \ \Lambda^2_n=\alpha^2\frac{\o_1}{k_{1n}},\ \ k_{1n}^2=k^2+\alpha^2
\label{vlk}
\eeq
and
\beq
E_{e0}=\frac{1}{2}\sum_n k_{1n},\ \ E_{X0}=\frac{1}{2}\sum_n \o_1,\ \
 E_{Y0}=\frac{1}{2}\sum_n \io \o.
\label{e0}
\eeq


\subsection{Two-stage Fano diagonalization}
In the approach of HB one first Fano-diagonalizes the "matter" Hamiltonian
$H_Y+H_{XY}$ with a real function $V(\o)$. The consistency relation is
automatically satisfied, since
\beq
\int_0^\infty \frac{d\o}{\o}V^2(\o)=\frac{1}{\o_1}\int_0^\infty d\o v^2(\o)<\o_1,
\label{consist1}
\eeq
which is fulfilled due to the definition of $\o_1$, Eq. (\ref{omn2}).
After this first step the total Hamiltonian acquires the form
\beq
H_1=\frac{1}{2}\sum_n\Big[k_{1n}\{a^\dagger_n,a_n\}+
\int_0^\infty d\o \o\{B^\dagger_{n\o},B_{n\o}\}+
i\Lambda_n[a^\dagger_n+a_n][b^\dagger_n-b_n]\Big].
\eeq
To transform it to the standard form (\ref{hgen}) we have to express $b$ in terms of
$B_\o$:
\beq
b_n=\int_0^\infty d\o\Big[ \alpha_0^*(\o)B_{n\o}-\beta_0(\o)B^\dagger_{n\o}
\Big]
\eeq
and its complex conjugate.
So
\[
b^\dagger_n-b_n=\int_0^\infty d\o\Big[B_{n\o}^\dagger
\Big(\alpha_0(\o)+\beta_0(\o)\Big)-c.c\Big]
\]
and the Hamiltonian $H_1$ takes the form (\ref{hgen}) with
\beq
V_n(\o)\to V_{1n}(\o)=i\Lambda_n\Big(\alpha_0(\o)+\beta_0(\o)\Big)
=-i\Lambda_n\o V(\o)Q^*(\o),
\eeq
where $Q(\o)$ is the "propagator" for the field $X$ in the medium
\[Q(\o)=\frac{1}{\o_0^2-\o^2\sigma(\o)}\]
and $\sigma$ playing the role of the "dielectric constant"
\[\sigma(\o)=1+\frac{\o_1}{2\o}
\io'\frac{d\o'}{\o'}\frac{|V(\o')|^2}{\o'-\o-i0}.
\]
Note that $k_{1n}|V_1(k,\o)|^2$ does not depend on $k_{1n}$.
One can check that
\[\io\frac{d\o}{\o}|V_{1n}(\o)|^2=\frac{\alpha^2}
{k_{1n}}<k_{1n}\]
and the consistency condition for the second diagonalization is fulfilled.

To finally diagonalize the Hamiltonian
we define operators
\beq
C_{n\o}=\xi_{0n}(\o)a_n+\eta_{0n}(\o) a^\dagger_n+
\int d\o'\Big(\xi_{1n}(\o,\o')B_{n\o'}+\eta_{1n}(\o,\o')B^\dagger_{n\o'}\Big).
\eeq
The resulting coefficients are found to be
\beq
\xi_{0n}=-\Big(\frac{\o+k_{1n}}{2}\Big)V_{1n}(\o)P_n^*(\o),\ \
\eta_{0n}(\o)=-\Big(\frac{\o-k_{1n}}{2}\Big)V_1(\o)P_n^*(\o)
\label{xieta}
\eeq
and
\beq
\xi_{1n}(\o,\o')=\delta(\o-\o')
-\frac{k_{1n}}{2}P_n^*(\o)\frac{V_{1n}^*(\o')V_{1n}(\o)}{\o-\o'-i0},
\label{xi1}
\eeq
\beq
\eta_{1n}(\o,\o')=-\frac{k_{1n}}{2}P_n^*(\o)
\frac{V_{1n}(\o')V_{1n}(\o)}{\o+\o'}.
\label{eta1}
\eeq
Here $P_n(\o)$ is the propagator of the electromagnetic field (\ref{prop1})
with the dielectric constant given by (\ref{ep}) with the
substitution $V\to V_1$.
\beq
\epsilon(\o)=1+\frac{k_{1n}}{2\o}
\int_{-\infty}^{\infty}\frac{d\o'}{\o'}\frac{|V_{1n}^2(\o')|}
{\o'-\o-i0}
\eeq
(as mentioned it does not depend on $k$).

The initial operators can be expressed via $C_\o(\bk)$.
Similarly to (\ref{oldnew})  we have
\beq
a_n=\int_0^\infty d\o\Big[ \xi_{0n}^*(\o)C_{n\o}-\eta_{0n}\o)
C^\dagger_{n\o}\Big]
\label{inv3}
\eeq
and expressing $B$ via $C$ we find
\beq
b_n=
\io\Big(\mu_{0n}^*(k,\o)C_{n\o}-\nu_{0n}(\o)C_{n\o}^\dagger\Big),
\label{inv4}
\eeq
where
\beq
\mu_{0n}(\o)=\io'\Big(\alpha_{0n}(\o')\xi_{1n}(\o,\o')+\beta_{0n}^*(\o')
\eta_{1n}(\o,\o')\Big),
\label{mu0n}
\eeq
\beq
\nu_{0n}(\o)=\io'\Big(\alpha_{0n}^*(\o')\eta_{1n}(\o,\o')+
\beta_{0n}(\o')\xi_{1n}(\o,\o')\Big)
\label{nu0n}
\eeq
and furthermore
\beq
b_{n\o}
=\io'\Big(\mu_{1n}^*(\o',\o)C_{n\o'}-\nu_{1n}(\o',\o)C_{n\o'}^\dagger\Big),
\label{inv5}
\eeq
where
\beq
\mu_{1n}(\o',\o)=\io"\Big(\alpha_1(\o",\o)\xi_{1n}(\o',\o")+
\beta_1^*(\o",\o)\eta_{1n}(\o',\o")\Big),
\label{mu1n}
\eeq
\beq
\nu_{1n}(\o',\o)=\io"\Big(\alpha_1^*(\o",\o)\eta_{1n}(\o',\o")+
\beta_1(\o",\o)\xi_{1n}(\o',\o")\Big).
\label{nu1n}
\eeq


\subsection{The ground state energy}
Expressing operators $a$, $b$ and $b_\o$ via $C_\o$ according to
Eqs. (\ref{inv3}), (\ref{inv4}) and (\ref{inv5}) and averaging in the
ground state with $C_{n\o}|0>=0$ find
\beq
<H_e>=\sum_nk_{1n}\io|\eta_{0n}|^2,
\label{he2}
\eeq
\beq
<H_X>=\sum_n\io \o_1|\nu_{0n}(\o)|^2,
\label{hx2}
\eeq
\beq
<H_Y>=\sum_n\io \o\io'|\nu_{1n}(\o',\o)|^2.
\label{hy2}
\eeq

Passing to the interaction terms
\beq
<H_{XY}>=\frac{1}{2}\sum_n\io V(\o)\io'
\Big(\mu_{0n}^*(\o')-\nu_{0n}^*(\o')\Big)\Big(\mu_{1n}(\o',\o)-\nu_{0n}(\o',\o)\Big)
\label{hxy2}
\eeq
and
\beq
<H_{eX}>=\frac{1}{2}i\sum_n\Lambda_n\io\Big(\xi_{0n}^*(\o)-\eta_{0n}^*(\o)\Big)
\Big(\mu_{0n}(\o)+\nu_{0n}(\o)\Big).
\label{hex2}
\eeq

Using relations between the coefficients we can rewrite the two
contributions from the interaction as
\beq
<H_{XY}=
\frac{1}{2}\sum_n\io V(\o)\io'
\Big[\nu_{1n}^*(\o',\o)\Big(\nu_{0n}(\o')-\mu_{0n}(\o')\Big)+c.c.\Big]
\label{hxy3}
\eeq
and
\beq
<H_{eX}>=
-\frac{1}{2}i\sum_n\Lambda_n\io\Big[\eta_{0n}^*(\o)\Big(\mu_{0n}(\o)+
\nu_{0n}(\o)\Big)-c.c\Big].
\label{hex3}
\eeq

Some remarks on the final formulas for calculations can be found in
the Appendix.

\section{Comparison of models with direct and indirect
interaction with the medium}
\subsection{Generalities}
Our central aim is to study whether the Casimir energy
depends only on the dielectric constant $\ep(\o)$ and
in this manner is independent of the model assumed for absorption
 or it depends on this model so that with
the same $\ep(\o)$ one gets different results for different
models for absorption. We have seen that in the simple model
studied in Sections 2. and 3. with the direct interaction (D model)
with the absorbing medium the Casimir energy is  expressed
entirely by  $\ep(\o)$.
The same is true for its simple generalizations considered in
Section 4. It remains to study the model introduced in Sections
5. -- 7. in which the electromagnetic field interacts indirectly
with the medium, via the atomic oscillator immersed in the medium
(HB model). Expression for the Casimir energy
in the latter model are far from being transparent and their inspection
does not allow to understand if the Casimir energy as before depends
entirely on the dielectic constant or it is not true. So we recur to
numerical check. We select parameters of the two models to give
the same dielectric constant and then calculate the resulting
Casimir energies.

A few words on the technique of the calculation.
The normalization of the Casimir energy, as mentioned, requires
subtracting the energy in absence of metallic plates, which implies
calculating the difference
\[ \sum_n-\int_0^\infty dn,\]
where in both terms $k_n=\pi n/a$. In this subtraction all terms
independent of $k_n$ do not contribute. Both the sum and the integral
contain integration over frequency $\o$. In the electromagnetic part
there are terms  divergent at $\o\to\infty$. To ensure convergence
it is convenient to separate ftom this part its value in absence of
the dielectric, which is the standard Casimir energy.
\[ E_{cas}^{(0)}=-\frac{\pi}{24a}.\]
Then after rotation to imaginary frequencies the energy
in model D changes to
\beq
E_0=\frac{1}{2\pi}\sum_n\int_0^\infty d\xi
\Big\{\Big(k_n^2-\xi^2\Big)\Big(P_n(i\xi)-P_n^{(0)}(i\xi)\Big)
-\xi^2\Big(\frac{d}{d\o}[\o(\ep(\o)-1)]\Big)_{\o=i\xi}P_n(i\xi)\Big\}.
\label{ce02}
\eeq
where $P_n^{(0)}(\o)=1/(k_n^2-\o^2-i0)$ is the free propagator.
In model HB only the electromagnetic part   is transformed
into
\beq
<H_e>_I=\frac{1}{2}\sum_n(k_{1n}-k_n)+
\frac{1}{2\pi}\,\sum_n\int_0^\infty d\xi
\Big[(k_n^2-xi^2)\Big(P_n(i\xi)-P_n^{(0)}(i\xi)\Big)+\alpha^2P_n(i\xi)\Big].
\eeq
The first term appears because $E_{e0}$ in (\ref{e0}) depends on the
interaction.

\subsection{Numerical calculations}
Integration over frequencies $\o$ cannot be efficiently done without
rotating the contour to pass along the imaginary axis, since the propagator
has a resonant behavior on the real axis. This rotation is trivial in model
D and for the electromagnetic part in model HB but not so trivial for
other terms in model HB, which contain factors with singularities in the
first quadrant of the complex $\o$-plane. In fact both the propagator and
the dielectric constant are regular in the upper half plane, as it should be.
But expression for the energy also depend on the complex conjugated quantities,
which are regular in the lower half plane but are allowed to have singularities
in the upper half plane. Also in any case one would like to
have analytic expressions for all terms in both models which allow to do
the analytic continuation constructively. So we have chosen a particular
form of the interaction $v(\o)$ in model HB which allows to
find all expressions including the dielectric constant
$\ep(\o)$ in the analytic form. This allowed to constructively do the
continuation to imaginary frquencies in both models.

\begin{figure}[]
\leavevmode \centering{\epsfysize=0.3\textheight\epsfbox{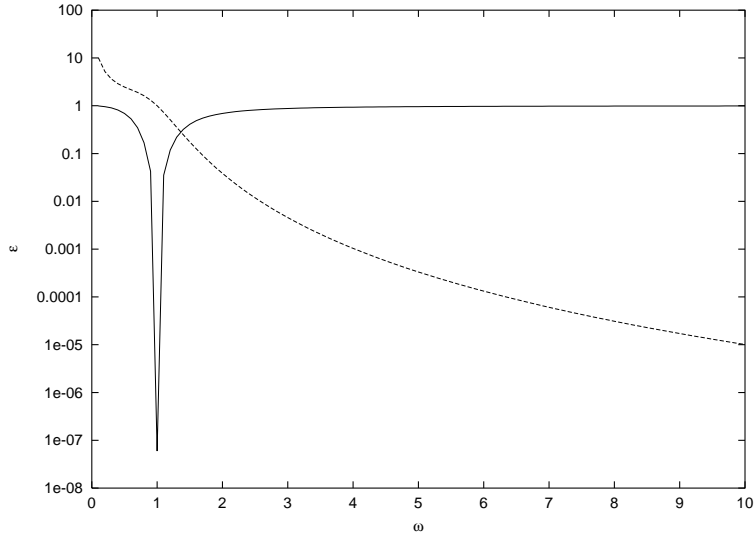}}
\caption{Real(upper curve for large $\o$) and imaginary parts
of $\ep(\o)-1$}
\label{fig1}
\end{figure}

Our choice is
\beq
v(\o)=\frac{g^2}{\o^2+m^2},\ \ g^2=\frac{1}{2\pi},\ \ \o_0=0.
\label{choice}
\eeq
With $m$ the only dimensionful parameter, the additional Casimir energy
 due to the interaction with the dielectric is
\[E_{cas}^{(1)}(a)=me_{cas}^{(1)}(ma).\]
In the following we put $m=1$.
With  choice (\ref{choice}) we find the dielectric constant
\beq
\ep(\o)=1-\alpha^2\frac{\o+i}{\o(\o^2-1+i\o)}
\eeq
We do the calculation for the interaction parameter $\alpha=1$.
For illustration we show real and imaginary parts of $\ep(\o)-1$ in this
case in Fig. \ref{fig1}.
In model HB we also meet quantities
$
k_{1n}|V_{1n}(\o)|^2=\alpha^2g^2\o/(\o^4-\o^2+1)
$
and $\epsilon^*(\o)$ which have a pole in the first quadrant
at $\o=\o_P=e^{1\pi/6}$. As mentioned they come from complex conjugated
terms, which appear in the expressions for the energy, in contrast to the
D model.
So in analytic continuation one had to take
into account residues at this pole.

\begin{figure}[]
\leavevmode \centering{\epsfysize=0.3\textheight\epsfbox{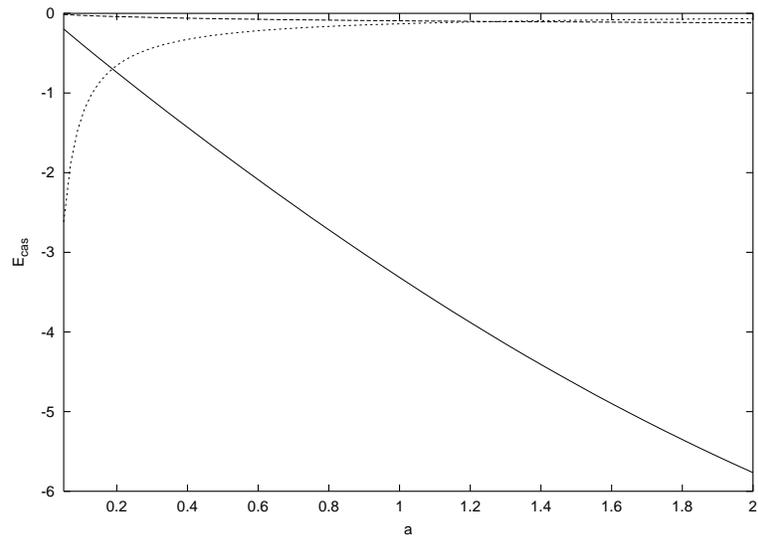}}
\caption{The additional Casimir energy in the dielectric in Models D
(middle curve at large $a$) and HB (lower curve at large $\o$).
The upper curve at large $\o$ shows the Casimir energy in the vacuum}
\label{fig2}
\end{figure}

\begin{figure}[]
\leavevmode \centering{\epsfysize=0.3\textheight\epsfbox{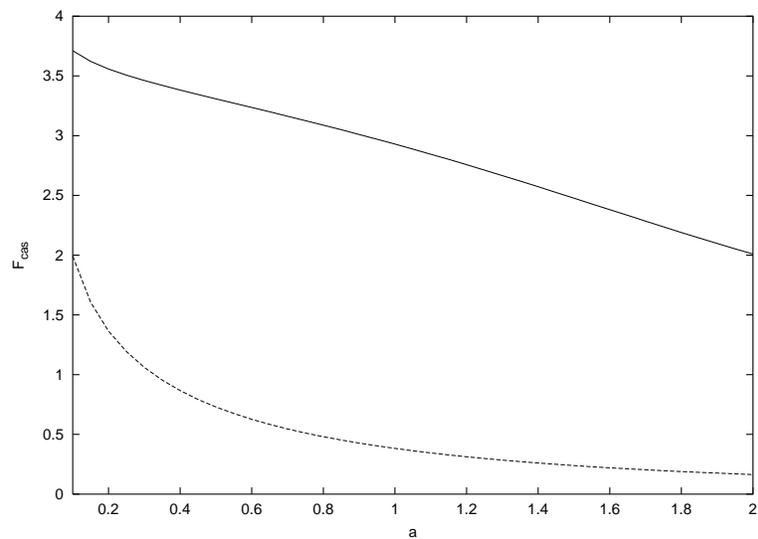}}
\caption{The additional Casimir force in the dielectric in Models D
(mutiplied by 10, lower curve) and HB (upper curve)}
\label{fig3}
\end{figure}

Our results for $E_{cas}^{(1)}(a)$ and the force $F_{cas}^{(1)}(a)$
are shown in Figs. \ref{fig2}and \ref{fig3} respectively.

Inspection of them clearly shows that the Casimir energy and force
do depend on the way the electromagnetic field interacts with the absorbing
medium. With indirect interaction both turn out to be considerably larger.
In model D at large $a>>1$  the additional energy tends to a finite
negative value $E_{cas}^{(1)}(a)\to -0.1618$
and the force  falls exponentially
$F_{cas}^{(1)}(a)\sim\exp(-0.055a)$. In contrast
in model IN  $E_{cas}^{(1)}(a)\sim -8\ln a$ and the force falls quite slowly
$F_{cas}^{(1)}(a)\sim 6/a$.
As to the absolute values of the additional contribution from the dielectric,
they depend on the values chosen for the parameters $\alpha$ and $m$.
So the curve describing the vacuum energy in Fig.
\ref{fig2} serves only for illustrative purpose.

\section{Conclusions}
A consistent quantization of the electromagnetic field in the
dielectric in the microscopic treatment remains an important
problem with far-reaching impact. It should give the answer to the
macroscopic treatment of the same problem, in which all influence
of the medium is entirely contained in the complex dielectric constant.
If this is indeed so then different microscopic models for dispersion
and absorption leading to the same dielectric constant
should give the same Casimir energy. Our study of two essentially different
models for the interaction with the medium, one in which the electromagnetic
field interacts ditectly with the  absorbing medium and the other via
the intermediary (atom) has shown that the answer is negative. With the
same dielectric constant the models give different Casimir energies and
forces. This means that the interaction with the dielectric in the
microscopic approach is not entirely encoded in the dielectric constant,
so that the Casimir energy has an extra dependence on the interaction with
the dielectric, which, in principle, can be studied experimentally.

Our conclusions are based on the study of very particular
(although well-known) microscopic models of absorption and moreover on
a specific parametrization of the interaction. However this is sufficient
to demonstrate the above-mentioned conclusion. In fact our formulas, in
principle, allow to calculate the Casimir energy for arbitrary
forms of interaction within the studied models. However realistic
calculations in model HB with indirect interaction with the medium seem
rather difficult. So at present we cannot find the Casimir energy in
model HB with different forms of interaction and so check whether the
found energy depends on these forms or is a characteristic of the model
as such.

\section{Acknowledgments}
This study has been supported by grants from SPbGU,
themes 11.0.59.2010,11.38.660.2013 and RFFI 12-02-00356/12.

\section{Comments on the final formulas for the calculation}
There exist certain relations between the coefficients introduced in
Section 5.2 which
guarantee fulfilment of commutation relations between initial operators.
Expressing
\beq
[a_n,b_{n'}]=[a_n,b_{n'}^\dagger]=0
\eeq
in terms of $C_\o$ one gets
\[\io\Big(-\xi_{0n}^*(\o)\nu_{0n}(\o)
+\eta_{0n}(\o)\mu_{0n}^*(\o)\Big)=0
\]
and
\[
\io\Big(\xi_{0n}^*(\o)\mu_{0n}(\o)
-\eta_{0n}(\o)\nu_{0n}^*(\o)\Big)=0.
\]

Similarly from
\beq
[b_n,b_{n'\o}]=[b_n,b_{n'\o}^\dagger]=0
\eeq
one finds
\[
\io'\Big(-\mu_{0n}^*(\o')\nu_{1n}(\o',\o)+
\nu_{0n}(\o')\mu_{1n}^*(\o',\o)\Big)=0
\]
and
\[
\io'\Big(\mu_{0n}^*(\o')\mu_{1n}(\o',\o)-
\nu_{0n}(\o')\nu_{1n}^*(\o',\o)\Big)=0.
\]
These relations can be used to simplify some expressions
for the ground state energy.

For particular cotributions to the energy we have the followingcomments.

{\bf 1.} $<H_e>$

Using (\ref{xieta}) we find
\beq
|\eta_{0n}(\o)|^2=\frac{1}{4}(\o-k_{1n})^2|V_{1n}|^2|P_n(\o)|^2.
\eeq
Since
\[ {\rm Im}\,P_n(\o)=\o^2{\rm Im}\,\epsilon(\o)|P_n(\o)|^2=
\pi\frac{k_{1n}}{2}|V_{1n}|^2|P_n(\o)|^2,\]
we obtain
\beq
<H_e>=\frac{1}{2\pi}{\rm Im}\,\sum_n\io (\o-k_{1n})^2P_n(\o).
\eeq
The integral admits rotation to the positive imaginary axis.

{\bf 2.}{$<H_X>$}

Coefficient $\nu_{0n}$ is given by Eq. (\ref{nu0n}) with
$\alpha_{0}$ and $\beta_0$ are given by (\ref{albe}).
and
coefficients $\xi_{1n}(\o,\o')$ and $\eta_{1n}(\o,\o')$ are given by formulas
(\ref{xi1}) and (\ref{eta1})
So
\[
\nu_{0n}=\Big(\frac{\o-\o_1}{2}\Big)\frac{V_{1n}(\o)}{i\Lambda_n\o}+
i\frac{k_{1n}}{4\Lambda_n}P_n^*(\o)V_{1n}(\o)J_1(\o),
\]
where
\beq
J_1(\o)=\io'\frac{|V_{1n}(\o')|^2}{\o'}\Big(\frac{\o'-\o_1}{\o-\o'-i0}
-\frac{\o'+\o_1}{\o+\o'}\Big)
=(\o-\o_1)J-2\io\frac{|V_{1n}(\o)|^2}{\o}
\eeq
with
\[
J=\io'\frac{|V_{1n}(\o')|^2}{\o'}
\Big(\frac{1}{\o-\o'-i0}+\frac{1}{\o+\o'}\Big)=
\int_{-\infty}^{+\infty}\frac{|V_{1n}(\o')|^2}{\o'}\frac{1}{\o-\o'-i0}
\]\beq=
\frac{2\o}{k_{1n}}.
\Big(1-\epsilon^*(\o)\Big).
\label{intj}
\eeq
The  second integral in $J_1$ is equal to $\alpha^2/k_{1n}$
So
\beq
J_1=
\frac{2}{k_{1n}}\Big[\o (\o-\o_1)\Big(1-\epsilon^*(\o)\Big)-\alpha^2\Big].
\eeq

This leads to our final result
\beq
\nu_{0n}(\o)=i\frac{V_{1n}(\o)}{2\Lambda_n\o}
\Big[\o_1-\o-(\o_1-\o)P_n^*(\o)\o^2\Big(1-\epsilon^*(\o)\Big)-
\alpha^2\o P_n^*(\o)\Big].
\label{nu0nf}
\eeq


{\bf 3.} $<H_Y>$

Coefficient $\nu_{1n}$ is given by (\ref{nu1n}). Coefficients
$\alpha_1$ and $\beta_1$ are given by (\ref{al1}) and (\ref{be1}).
Coefficients $\xi_{1n}$ and $\eta_{1n}$ are given by (\ref{xi1})
and (\ref{eta1}).
As a result we find $\nu_{1n}$ as a sum of three terms
\[
\nu_{1n}(\o',\o)=
-\frac{k_{1n}}{2}P_n^*(\o')
\frac{V_{1n}(\o)V_{1n}(\o')}{\o'+\o}
-i\frac{\o_1}{2\Lambda_n \o'}
\frac{V(\o)V_{1n}(\o')}{\o'+\o}\]\beq
+i\frac{k_{1n}}{4\alpha}\sqrt{k_{1n}\o_1}P_n^*(\o')V(\o)V_{1n}(\o')J_2,
\label{nu1nf}
\eeq
where
\[
J_2=\io"\frac{|V_{1n}(\o")|^2}{\o"}
\Big(-\frac{1}{\o'+\o"}\,\frac{1}{\o"-\o+i0}+
\frac{1}{\o+\o"}\,\frac{1}{\o'-\o"-i0}\Big)\]\[
=\frac{1}{\o+\o'}\io"\frac{|V_{1n}(\o")|^2}{\o"}
\Big(
\frac{1}{\o-\o"-i0}+\frac{1}{\o+\o"}+\frac{1}{\o'-\o"-i0}+\frac{1}{\o'+\o"}
\Big)
\]
\[=
\frac{1}{\o+\o'}\Big(J(\o)+J(\o')\Big)=
\frac{2}{k_{1n}}\frac{\o\Big(1-\epsilon^*(\o)\Big)+
\o'\Big(1-\epsilon^*(\o')\Big)}{\o+\o'}.
\]

{\bf 4.} $<H_{XY}>$

Coefficients $\nu_{0n}$ and $\nu_{1n}$ are already given
by(\ref{nu0nf}) and (\ref{nu1nf}). Coefficient $\mu_{0n}$
 differs from $\nu_{0n}$ by the change $\alpha_0\leftrightarrow\beta_0$,
that is $\o_1\to -\o_1$
So
\beq
\mu_{0n}(\o)=i\frac{V_{1n}(\o)}{2\Lambda_n\o}
\Big[-\o_1-\o+(\o_1+\o)P_n^*(\o)\o^2\Big(\epsilon(\o)-1\Big)-
\alpha^2\o P_n^*(\o)\Big].
\label{mu0nf}
\eeq

{\bf 5.}{$<H_{eX}>$}

Here all coefficients are already known.


\begin{thebibliography}{99}
%
\bibitem{hutbar} B.Huttner and S.M.Barnett,
Phys. Rev. {\bf A 46} (1992) 4306.
%
\bibitem{mochan} W.L.Mochan and C.Villarreal,
New J.Phys.{\bf 8} (2006) 241.
%
\bibitem{lombardo} F.C.Lombardo, F.D.Mazzitelli and A.E. Rubio Lopez,
Phys. Rev {\bf A 84} (2012) 053517
%
\bibitem{ram1} F.S.S.Rosa, D.A.R.Alvit and P.W.Milloni,
Phys. Rev. {\bf A 81} (2010) 033812.
%
\bibitem{interavia} F.Interavia and R.Behunin, Phys. Rev. {\bf A 86}
(2012) 062517.
%
\bibitem{ram2}  F.S.S.Rosa, D.A.R.Alvit and P.W.Milloni, arxiv: 0912.0279
%
\bibitem{phil} T.Philbin, New J.Phys. {\bf 13}(2011) 063026.
%
\bibitem{lifshitz} E.M.Lifshitz, Zh. Eksp. Teor. Fiz. {\bf 29} (1955) 94.
%
%
\end{thebibliography}
\end{document}